\documentclass[12pt]{article}
\usepackage[dvips]{graphicx}
\usepackage{amsmath,amssymb,amsfonts}

\begin{document}

\title{Reinforcement of $K^0_{S}$ regeneration in the model with Hermitian Hamiltonian}
\author{V.I. Nazaruk\\
Institute for Nuclear Research of RAS, 60th October\\
Anniversary Prospect 7a, 117312 Moscow, Russia.*}

\date{}
\maketitle
\bigskip

\begin{abstract}
The model of the $K^0_{S}$ regeneration with Hermitian Hamiltonian is proposed. An increase of regeneration takes place.
\end{abstract}

\vspace{5mm}

\vspace{5mm}

\vspace{1cm}

*E-mail: nazaruk@inr.ru

\newpage
\setcounter{equation}{0}

\section{Introduction}
In standard calculations of the $K^0_{S}$ regeneration [1-5] non-coupled equations of motion are used instead of the coupled ones. The off-diagonal mass was omitted. We note that in [5] a correct scheme of calculation is described, but the result (see Eqs. (7.83)- (7.89) of Ref. [5]) corresponds to non-coupled equations (see [6] for more details and Sect. 2).

This is the principal error which entails a qualitative disagreement in the results. In this regard a model based on the exact solution of coupled equations of motion has been proposed [6, 7]. The basic element of this model is an optical potential. The next question is the applicability of the optical potential in the system of the coupled equations. It was shown in [8] that for a system of equations in the strong absorption region the optical potential is inapplicable. 

In this paper we present a simple model of the direct calculation with a Hermitian Hamiltonian. The optical potential is not used.  (For the $n\bar{n}$ transitions in the medium the calculations beyond the potential model were presented in [9-11].)
 
Let us denote models based on the optical potential as potential models. 

\section {Potential models}
The fundamental difference between our and previous calculations lies in the process model. For the previous old calculations [2] the starting equations are (see Eqs. (3) from [2]):
\begin{eqnarray}
(\partial_x-ink)K^0=0,\nonumber\\
(\partial_x-in'k)\bar{K}^0=0,
\end{eqnarray}
where $n$ and $n'$ are the indexes of refraction for $K^0$ and $\bar{K}^0$, 
respectively. In notations of Ref. [2] $K^0=\alpha $ and $\bar{K}^0=\alpha '$,
$K^0_{S}=\alpha _1$ and $K^0_{L}=\alpha _2$. In above-given Eq. (1) we substitute $K^0=(\alpha _1+ i\alpha _2)/ \sqrt{2}$, $\bar{K}^0=(\alpha _1-i\alpha _2)/ \sqrt{2}$ and include the effect of the weak interactions as in [2]. We obtain Eq. (5) and result (6) from Ref. [2].

However, the coupled equations should be used:
\begin{eqnarray}
(i\partial_t-M)K^0=\epsilon \bar{K}^0,\nonumber\\
(i\partial_t-(M+V))\bar{K}^0=\epsilon K^0,
\end{eqnarray}
where
\begin{eqnarray}
M=m_{K^0}+U_{K^0}-i\Gamma _{K^0}^d/2,\nonumber\\
V=(m_{\bar{K}^0}-m_{K^0})+(U_{\bar{K}^0}-U_{K^0})-(i\Gamma _{\bar{K}^0}^d/2-i\Gamma _{K^0}^d/2).
\end{eqnarray}
Here $\epsilon =(m_L-m_S)/2=\Delta m/2$ is a small parameter, $U_{K^0}$ and $U_{\bar{K}^0}$ are the potentials of $K^0$ and ${\bar{K}^0}$, $\Gamma _{K^0}^d$ and 
$\Gamma _{\bar{K}^0}^d$ are the decay widths of $K^0$ and ${\bar{K}^0}$, respectively.

So the starting equations (3) from [2] are non-coupled. There is no off-diagonal mass $\epsilon =(m_L-m_S)/2$. This is a fundamental defect. The non-coupled equations exist only for the stationary states and don't exist for $K^0$ and $\bar{K}^0$. Equations (2) given above 
should be considered, not (1). 

In this connection the model based on the exact solution of coupled equations of motion was proposed [6,7]. However, it was shown in [8] that for a system of equations in the strong absorption region the optical potential is inapplicable. Bellow we present the calculations based on the $S$-matrix approach. The optical potential is not used.

\section{$S$-matrix approach}
We consider the process
\begin{equation}
K^0_{L}\rightarrow K^0_{S}\rightarrow \pi \pi.
\end{equation}
For definiteness, the decay into $2\pi $ is registered [12]. Let $K^0_{L}$ falls onto the plate at $t=0$. Notations of Ref. [5] are used.
Since $K^0N$- and $\bar{K}^0N$-intereactions are known, we go into basis $(K^0,\bar{K}^0)$ using the relation 
\begin{equation}
\left.\mid\!K^0_L\right>=\left.(\mid\!K^0\right>+\left.\mid\!\bar{K}^0\right>)/\sqrt{2}.
\end{equation}
Here $\left.\mid\!K^0\right>$, $\left.\mid\!\bar{K}^0\right>$ and $\left.\mid\!K^0_L\right>$ are the states of $K^0$, $\bar{K}^0$ and $K^0_L$, respectively. 

At the time $t$ we have [5]
\begin{equation}
\left.\mid\!K^0_L(t)\right>=\left.(\mid\!K^0(t)\right>+\left.\mid\!\bar{K}^0(t)\right>)/\sqrt{2}=\left.(\mid\!K^0\right>e^{-\lambda _{K^0}t} +\left.\mid\!\bar{K}^0\right>e^{-\lambda _{\bar{K}^0}t})/\sqrt{2}.
\end{equation}
Here
\begin{eqnarray}
\lambda _{K^0}=im_{K^0}+\Gamma _{K^0}/2,\nonumber\\
\lambda _{\bar{K}^0}=im_{\bar{K}^0}+\Gamma _{\bar{K}^0}/2,
\end{eqnarray}
where
\begin{eqnarray}
\Gamma _{K^0}=\Gamma _{K^0}^a+\Gamma _{K^0}^d,\nonumber\\
\Gamma _{\bar{K}^0}=\Gamma _{\bar{K}^0}^a+\Gamma _{\bar{K}^0}^d.
\end{eqnarray}
Here $\Gamma _{K^0}^a$ and $\Gamma _{K^0}^d$ are the widths of absorption and decay of $K^0$; $\Gamma _{\bar{K}^0}^a$ and $\Gamma _{\bar{K}^0}^d$ are the widths of absorption and decay of $\bar{K}^0$, respectively. 

The wave function of $K^0$ is
\begin{equation}
K^0(x)=\Omega ^{-1/2}\exp (-ipx),
\end{equation}
where $p=(E,{\bf p})$ is the 4-momentum of $K^0$; $E={\bf p}^2/2m+U_{K^0}$. To draw an analogy with the well-studied 
$n\bar{n}$ transitions we consider a non-relativistic problem. The potential $U_{K^0}$ is real. The unperturbed and interaction Hamiltonians are
\begin{eqnarray}
H_0=-\nabla^2/2m+ U_{K^0},\nonumber\\
H_I=H_{K^0\bar{K}^0}+H_W+V,\nonumber\\
V=U_{\bar{K}^0}-U_{K^0},\nonumber\\
H_{K^0\bar{K}^0}=\int d^3x(\epsilon \bar{\Psi }_{\bar{K}^0}(x)\Psi _{K^0}(x)+H.c.).
\end{eqnarray}
Here $\epsilon =(m_L-m_S)/2$ [6]; $U_{K^0}$ and $U_{\bar{K}^0}$ are the real potentials of $K^0$ and $\bar{K}^0$,
$H_{K^0\bar{K}^0}$ and $H_W$ are the Hamiltonians of the $K^0\bar{K}^0$
conversion and decay of the $K$-mesons, respectively; $\bar{\Psi }_{\bar{K}^0}$ and
$\Psi _{K^0}$ are the fields of $\bar{K}^0$ and $K^0$, respectively.

For the amplitide of process (4) we have
\begin{eqnarray}
M(K_L\rightarrow K_S\rightarrow \pi \pi )=<\pi \pi \mid H_I\mid\!K_L(t)>=\nonumber\\
\frac{1}{\sqrt{2}}[<\pi \pi \mid H_I\mid\!K^0>e^{-\lambda _{K^0}t}+<\pi \pi \mid H_I\mid\!\bar{K}^0>e^{-\lambda _{\bar{K}^0}t}]. 
\end{eqnarray}

The matrix element $<\pi \pi \mid H_I\mid\!K^0>$ is calculated by means of the perturbation theory. In the lowest order in $H_{K^0\bar{K}^0}$ we have
\begin{eqnarray}
<\pi \pi \mid H_I\mid\!K^0>=<\pi \pi \mid H_W\mid\!K^0>+\epsilon G(\bar{K}^0)<\pi \pi \mid H_W\mid\!\bar{K}^0>,\nonumber\\
G(\bar{K}^0)=-1/V
\end{eqnarray}
(see Fig. 1). Here $G(\bar{K}^0)$ is the propagator of $\bar{K}^0$; $<\pi \pi \mid H_W\mid\!K^0>$ and Fig. 1a correspond to the zero 
order in $H_{K^0\bar{K}^0}$. They describe the direct decay $K^0\rightarrow \pi \pi $. Figure 1b and the term $\epsilon G(\bar{K}^0)<\pi \pi \mid H_W\mid\!\bar{K}^0>$ correspond to the first order in $H_{K^0\bar{K}^0}$ and all the orders in $V$ and $H_W$.

\begin{figure}[!h]
%  \reflectbox{\includegraphics[height=.25\textheight]{golfer}}
  {\includegraphics[height=.2\textheight]{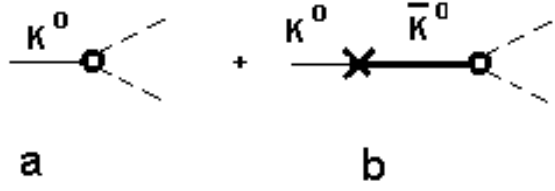}}
\caption{The decay $K^0\rightarrow 2\pi $ in the medium: (a) and (b) correspond to the
zero and first orders in $H_{K^0\bar{K}^0}$, respectively.} 
\end{figure}

The regeneration arises from the difference between the $K^0N$- and $\bar{K}^0N$-intereactions. It is induced by $H_{K^0\bar{K}^0}$ and so we hold only the second term in (12):
\begin{equation}
<\pi \pi \mid H_I\mid\!K^0>=\epsilon G(\bar{K}^0)<\pi \pi \mid H_W\mid\!\bar{K}^0>.
\end{equation}

Similarly, for the second term in (11) we obtain
\begin{eqnarray}
<\pi \pi \mid H_I\mid\!\bar{K}^0>=\epsilon G(K^0)<\pi \pi \mid H_W\mid\!K^0>,\nonumber\\
G(K^0)=1/V.
\end{eqnarray}

Substituting (13) and (14) into (11) we get
\begin{eqnarray}
M(K_L\rightarrow K_S\rightarrow \pi \pi )=\frac{1}{\sqrt{2}}[\epsilon G(\bar{K}^0)<\pi \pi \mid H_W\mid\!\bar{K}^0>e^{-\lambda _{K^0}t}+\nonumber\\
\epsilon G(K^0)<\pi \pi \mid H_W\mid\!K^0>e^{-\lambda _{\bar{K}^0}t}]. 
\end{eqnarray}

We return to $K_L,K_S$ representation.
\begin{eqnarray}
M(K_L\rightarrow K_S\rightarrow \pi \pi )=\frac{\epsilon }{2}<\pi \pi \mid H_W\mid\!K_S>[G(K^0)e^{-\lambda _{\bar{K}^0}t}-G(\bar{K}^0)e^{-\lambda _{K^0}t}]+F,\nonumber\\
F=\frac{\epsilon }{2}<\pi \pi \mid H_W\mid\!K_L>[G(\bar{K}^0)e^{-\lambda _{K^0}t}+G(K^0)e^{-\lambda _{\bar{K}^0}t}].
\end{eqnarray}

If $<\pi \pi \mid H_W\mid\!K_L>=0$, then we have
\begin{equation}
M(K_L\rightarrow K_S\rightarrow \pi \pi )=\frac{\epsilon }{2V}<\pi \pi \mid H_W\mid\!K_S>(e^{-\lambda _{K^0}t}+e^{-\lambda _{\bar{K}^0}t}). 
\end{equation}
We put $m_{K^0}=m_{\bar{K}^0}=m$, $\Gamma _{\bar{K}^0}^d=\Gamma _{K^0}^d=\Gamma ^d$. Now the process amplitude is as follows
\begin{equation}
M(K_L\rightarrow K_S\rightarrow \pi \pi )=\frac{\epsilon }{2V}<\pi \pi \mid H_W\mid\!K_S>(1+e^{-\Delta \Gamma t/2})e^{-i(im+\Gamma ^t/2)t}.
\end{equation}
where
\begin{eqnarray}
\Delta \Gamma =\Gamma _{\bar{K}^0}^a-\Gamma _{K^0}^a,\nonumber\\
\Gamma ^t=\Gamma _{K^0}^d+\Gamma _{K^0}^a.   
\end{eqnarray}
Finally
\begin{eqnarray}
\mid M(K_L\rightarrow K_S\rightarrow \pi \pi )\mid^2=J\mid <\pi \pi \mid H_W\mid\!K_S>\mid^2,\nonumber\\
J=\frac{\epsilon ^2}{4V^2}(1+e^{-\Delta \Gamma t/2})^2e^{-\Gamma ^tt}.
\end{eqnarray}

The width of process (4) is
\begin{eqnarray}
\Gamma (K_L\rightarrow K_S\rightarrow \pi \pi )=N\int d\Phi \mid M(K_L\rightarrow K_S\rightarrow \pi \pi )\mid^2=J\Gamma _d(K_S\rightarrow \pi \pi ),\nonumber\\
\Gamma _d(K_S\rightarrow \pi \pi )=N\int d\Phi \mid<\pi \pi \mid H_W\mid\!K_S>\mid^2, 
\end{eqnarray}
where $\Gamma _d(K_S\rightarrow \pi \pi )$ is the width of the decay $K_S\rightarrow \pi \pi $.

Let the probability of finding $\pi \pi $ be given by the exponential decay law. In the lowest order in $\Gamma (K_L\rightarrow K_S\rightarrow \pi \pi )$
\begin{equation}
W_S(K_L\rightarrow K_S\rightarrow \pi \pi )=1-e^{-\Gamma (K_L\rightarrow K_S\rightarrow \pi \pi )t}\approx \Gamma (K_L\rightarrow K_S\rightarrow \pi \pi )t.
\end{equation}
(By means of the values given below it is easy to verify that $\Gamma (K_L\rightarrow K_S\rightarrow \pi \pi )t\ll 1$.)

\begin{figure}[h]
%  \reflectbox{\includegraphics[height=.3\textheight]{golfer}}
  {\includegraphics[height=.25\textheight]{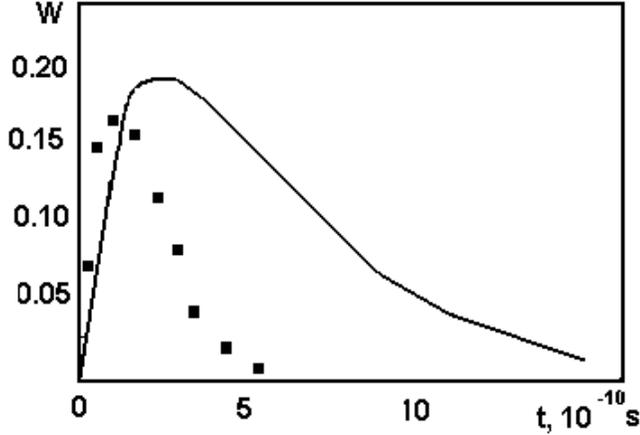}}
  \caption{ Probability of finding $K^0_S$.}
\end{figure}

In the standard calculations [1-5] $K^0_{S}$ in the final state is considered whereas in (22) a probability of finding $\pi \pi $ in the final state is given. To 
compare the results we modify our model. Let us assume that  
\begin{eqnarray}
W_S(K_L\rightarrow K_S\rightarrow \pi \pi )=W_S(K_L\rightarrow K_S)W,\nonumber\\
W=\Gamma _d(K_S\rightarrow \pi \pi )/\Gamma ^t,
\end{eqnarray}
where $W_S(K_L\rightarrow K_S)$ and $W$ are the probabilities of $K_LK_S$ transition and decay $K_S\rightarrow \pi \pi $, respecitively. Substituting (22), we obtain
\begin{equation}
W_S(K_L\rightarrow K_S)=J\Gamma ^tt.
\end{equation}

\section{Results and discussion}
For a copper absorber the probability of finding $K^0_{S}$ is shown in Fig. 2. The curve and squares depict the calculation performed using equation (24) and the old results given in [2], respectively. In our calculation we take $\sigma (K^0n)=\sigma (K^0p)=15$ mb [13], $V= \Delta \Gamma /2$. Like in Ref. [2], we use $\sigma (K^0N)=\frac{1}{3}\sigma (\bar{K}^0N)$.

The old results are given only for illustration. For the particle oscillations in absorbing matter there are two questions:

1) In the old calculations all the results have been obtained from non-coupled equations of motion. The term $H_{K^0\bar{K}^0}$ (the off-diagonal mass) was omitted; however it plays a leading part in oscillations. We would like to emphasize that all the results given in [2-5] follow from non-coupled equations mentioned above. For example, Eq. (7.90) from Ref. [5]. In this connection the model based on the exact solution of coupled equations of motion was proposed [6,7]. The comparison of the results of the present paper and Ref. [7] will be given in the next paper.

2) The applicability of optical potential in the system of coupled equations of motion. In the calculation given above the system of equations (optical potential) is not used.

The above-given calculation is free from both drawbacks. The main results of this paper are given by Eqs. (23) and (24). They are simple and transparent.

\end{document}